\documentclass{aa}
\usepackage[utf8]{inputenc}
\usepackage{amsmath}
\usepackage{amsfonts}
\usepackage{amssymb}
\usepackage{color,soul}
\usepackage{longtable}
\usepackage{graphicx}

\begin{document}


\title{Thermal fracturing on comets}
\subtitle{Applications to 67P/Churyumov--Gerasimenko}
\author{N.~Attree \inst{1} \and O.~Groussin \inst{1} \and L.~Jorda \inst{1} \and S.~Rodionov \inst{1} \and  A-T.~Auger \inst{1}
\and N.~Thomas \inst{2} \and Y.~Brouet \inst{2} \and O.~Poch \inst{2}
\and E.~K{\"u}hrt \inst{3} \and  M.~Knapmeyer \inst{3} \and F.~Preusker \inst{3} \and F.~Scholten \inst{3} \and J.~Knollenberg \inst{3} \and S.~Hviid \inst{3}
\and P.~Hartogh \inst{4}
}

\institute{Aix Marseille Univ, CNRS, LAM, Laboratoire d'Astrophysique de Marseille, Marseille, France, \email{Nicholas.Attree@lam.fr}
\and Physikalisches Institut, Universit{\"a}t Bern, Sidlerstrasse 5, 3012 Berne, Switzerland
\and Deutsches Zentrum f{\"u}r Luft-und Raumfahrt (DLR), Institut f{\"u}r Planetenforschung, Rutherfordstra{\ss}e 2, 12489 Berlin, Germany
\and Max-Planck-Institut f{\"u}r Sonnensystemforschung, Justus-von-Liebig-Weg 3, 37077 G{\"o}ttingen, Germany
}


\abstract{
We simulate the stresses induced by temperature changes in a putative hard layer near the surface of comet 67P/Churyumov--Gerasimenko with a thermo-viscoelastic model. Such a layer could be formed by the recondensation or sintering of water ice (and dust grains), as suggested by laboratory experiments and computer simulations, and would explain the high compressive strength encountered by experiments on board the Philae lander. Changes in temperature from seasonal insolation variation penetrate into the comet's surface to depths controlled by the thermal inertia, causing the material to expand and contract. Modelling this with a Maxwellian viscoelastic response on a spherical nucleus, we show that a hard, icy layer with similar properties to Martian permafrost will experience high stresses:  up to tens of MPa, which exceed its material strength (a few MPa), down to depths of centimetres to a metre. The stress distribution with latitude is confirmed qualitatively when taking into account the comet's complex shape but neglecting thermal inertia. Stress is found to be comparable to the material strength everywhere for sufficient thermal inertia ($\gtrsim50$ J m$^{-2}$ K$^{-1}$ s$^{-1/2}$) and ice content ($\gtrsim 45\%$ at the equator). In this case, stresses penetrate to a typical depth of $\sim0.25$ m, consistent with the detection of metre-scale thermal contraction crack polygons all over the comet. Thermal fracturing may be an important erosion process on cometary surfaces which breaks down material and weakens cliffs.}



\maketitle

\section{Introduction}

Fracturing is prevalent on many scales on  Comet 67P/Churyumov--Gerasimenko (hereafter  67P)  when observed by Rosetta's OSIRIS imaging instrument. Large fractures extending up to several hundred metres in length are obvious in several places on exposed consolidated terrains (see \citealp{Thomas15} and \citealp{ElMaarry15} for a detailed description of terrain types), such as in the lineaments on the Hathor cliffs or in the $\sim500$ m  feature near the neck \citep{ElMaarry2015}. On smaller scales (metres to decimetres), irregular fracture networks are seen globally, covering many of the consolidated and brittle terrain types, as well as many cliff faces. Fractured cliffs are commonly associated with mass wasting and landslides \citep{Pajola15}, and fracturing may represent an important erosion process. Pervasive, semi-regular networks of intersecting fractures have been likened to thermal contraction crack polygons, also seen on Earth and Mars \citep{ElMaarry2015, ElMaarry15, Auger15, Auger17}. Polygonal fractures and other patterns often overlay each other, implying different formation mechanisms or timeframes. Morphologically, most individual fractures resemble tensile, `opening' features \citep{ElMaarry2015}, but a population of fractures associated with the neck region have been identified as a possible shear zone \citep{Matonti}. Fractures are also present on the surface of the boulders  which litter the cometary surface, with examples down to a few metres in size covered in fractures that are from tens of centimetres to metres long \citep{Pajola15} and often surrounded by fragmented remains. An erosion sequence of chunks of consolidated cliff material into ever smaller boulders is implied;  \cite{Pajola15} have suggested a balance between this process and sublimation, leading to the observed steady-state distribution in boulder sizes. At the smallest scales, the Philae lander's CIVA camera saw tens of cm to sub-cm fractures covering the boulders and blocks making up the Abydos landing site \citep{Poulet16}. They are noted to be neither linear nor regularly polygonal, but with a morphology suggestive of a slow stress evolution.

Such a variety of fracture types likely have a range of formation mechanisms;  rotational, collisional, and thermal mechanisms have all been proposed. Aligned fracture systems in the neck may be associated with torques from rotation- or orbit-induced stresses or, alternatively, with stresses from a putative collision and merger of the two comet lobes \citep{ElMaarry2015, Matonti}. Desiccation stresses, from changes in volume caused by loss of volatiles, are important on Earth and may also play a role where outgassing is occurring on the comet. Other features, especially the polygonal networks, are more likely caused by thermal stresses, as evidenced by their uniformity over the nucleus and their tensile morphology, and supported by the large changes in temperature expected on cometary surfaces. \cite{ElMaarry2015} compute diurnal temperature changes of $\sim230$ K for 67P at perihelion and this is similar in magnitude to the Moon, where thermal weathering has been shown to be important \citep{Molaro15}. \cite{Lagoa2015} also invoke thermal weathering to explain activity on the comet starting before perihelion. Thermal stresses may well be the dominant mechanism for the formation of polygonal and irregular fractures (those not associated with the neck) on both  large  \citep{Thomas15,ElMaarry15} and small scales \citep{Poulet16} on 67P, and the remainder of this paper will focus on these processes.

Thermal stresses are induced by changes in temperature causing the expansion and contraction of material (in both uniform samples and mixtures with varying thermal expansion coefficients) leading to breakdown in a process known as thermal insolation weathering. If stresses exceed a material's tensile strength at any time, failure will occur immediately, leading to a thermal shock fracture. Alternatively, repeated stress cycles below the failure stress can lead to gradual weakening and the development of micro-fractures, leading to eventual failure in a process known as thermal fatigue \citep{HallThorn14, Delbo14}. Thermal fatigue follows existing material weaknesses, often leading to fracturing parallel to the surface (flaking), whereas thermal shock cuts across existing weaknesses, forming distinctive rectilinear or polygonal fracture networks. Links between the two are poorly understood, but it is unlikely that thermal fatigue enhances the probability of fractures developing by thermal shock \citep{HallThorn14}.

Thermal fatigue may be the dominant mechanism for the formation of regolith on asteroids \citep{Delbo14, Dombard2010}. When subjected to diurnal temperature variations of the  order of 100 K, rocks larger than a few centimetres are broken down by the gradual growth of fractures. Thermal fatigue may also be important on comets \citep{Lagoa2015}, but large temperature variations mean that immediate material breakdown by thermal shock may dominate. \cite{Kuehrt84} modelled Halley's comet as solid water ice and showed that seasonal thermal stresses at perihelion can reach tens of MPa, easily exceeding the tensile strength. These results were robust when extended from merely elastic stresses to a viscous description  \citep{TauberKuehrt87}, and similar when considering inhomogeneous bodies, showing that comets should be thermally fractured and that this may be important for erosion and outgassing.

On Earth and Mars, stresses exceeding a material's tensile strength over large, uniform areas lead to networks of fractures which intersect in polygonal patterns. Each fracture acts to dissipate stress to some distance around itself, depending on the material properties and fracture depth, thus determining the distance to the next fracture. Polygon size is therefore related to the material and thermal environment in the subsurface \citep{Lachenbruch62}. Thermal contraction crack polygons are seen in terrestrial glaciers (e.g.~\citealp{Marchant02}), bedrock (e.g.~\citealp{ElMaarry10}), and ice-bonded soils (e.g.~\citealp{Mellon97, Lachenbruch62}). \cite{Auger17} mapped polygons on the surface of 67P, finding them on both comet lobes, in the northern and southern hemispheres, and with no particular correlation with latitude or other features. Their morphology closely resembled terrestrial and Martian features, strongly suggesting thermal fracturing as the origin (see e.g.  their Figure 12). On Earth, polygon formation is enhanced by the presence of water ice (because of its large coefficient of thermal expansion and because it further opens and erodes  fractures by freeze-thaw weathering), but the presence of ice is not a prerequisite \citep{HallThorn14}. Water ice would appear to be required on 67P in order to bond the otherwise poorly consolidated nucleus material together to form a layer competent enough for high stresses to develop. \cite{Auger15, Auger17} suggest that a hard sintered layer within a few centimetres of the surface would be consistent with the observed polygons of a few metres in size. We discuss the evidence for such a hard layer in Section \ref{evidence}. In Section \ref{modelling}, we introduce a thermo-mechanical model of the cometary surface, with the aim of further quantifying the thermal stresses on 67P and their variations. The results of this model are discussed in Section \ref{results}, followed by the implications and conclusions in Section \ref{conclusion}.

\section{Evidence for a hard layer}
\label{evidence}

The presence of thermal contraction crack polygons implies a competent layer within centimetres or metres of the cometary surface. Additional evidence also supports this hypothesis and we discuss it now.

\subsection{Constraints from laboratory experiments}

Since the 1950s, several laboratory simulation experiments have been performed in order to understand the physical processes occurring on cometary nuclei. These experiments consist of producing centimetre-thick porous and fluffy samples made of particles of H$_{2}$O (and CO$_{2}$) ice and dust (silicates, organics, etc.) maintained in a cold thermal-vacuum chamber while irradiated with a Sun-simulator. The irradiation of these ice-dust samples triggers the sublimation of the ice, producing a dehydrated dust mantle at the surface. Underneath this mantle, a hardening of the remaining ice was observed in many experiments \citep{Kochan89, RatkeKochan, Seiferlin95, Thiel89, Poch2016a, Poch2016b}. This consolidation can be explained by the re-condensation of H$_{2}$O/CO$_{2}$ between the individual ice and dust particles and/or by sintering: both processes build bonds connecting the ice particles together and cause solidification of the sample. Measurements of the hardness of these ice-dust crusts, performed after the KOSI (Kometensimulation, DLR) experiments, yielded values of 0.15 to 5 MPa \citep{Grun93, Kochan89}. Recently, new simulation experiments performed at the Laboratory for Outflow Studies of Sublimating Materials (LOSSy) of the University of Bern showed that the hardening depends on the way the ice and the dust are initially mixed together: sublimation of an intimate mixture of ice and dust at the particle level (inter-mixture) leads to a higher solidification than a mixture where the dust particles are included in the grains of ice (intra-mixture) \citep{Poch2016b}. 

\subsection{Constraints from modelling}

\cite{Kossacki15} modelled the sintering of water ice grains in the upper layers of 67P using a thermophysical model of a mixture of ice and mineral grains under a dusty mantle (thickness of $0-16$ cm), subjected to solar heating. Sublimation and recondensation of the water ice reshaped the grains, enhancing their surface contact area and increasing the tensile and compressive strengths. Depending on the initial grain size, thickness of the dust layer, and insolation conditions, \cite{Kossacki15} found a sintered layer extending up to several metres below the dust mantle after a few orbital periods. Within this layer the strength was enhanced, relative to an unconsolidated mixture, by several orders of magnitude, so that it was comparable to a few times less than the $\sim$MPa strength of solid water ice. \cite{Kossacki15} use $1, 2.5$, and $10$ MPa tensile strengths for, respectively, pure poly-crystalline ice at $240-270$ K and at $100$ K, and an ice-rock mixture of roughly $20\%$ rock by mass at $100$ K. This hardening occurred at both the locations studied (the equator and the Agilkia landing site) and is expected elsewhere whenever the insolation is high enough (making it less efficient in shadowed areas) and the initial grain size sufficiently small (tens of microns in radius or smaller). 

\subsection{Constraints from the lander}

Two Philae instruments, SESAME-CASSE and MUPUS-PEN,  provide direct clues about the mechanical properties of the near-surface layer of the comet at the Abydos landing site. 
At Abydos, CASSE listened to the hammering of the MUPUS PEN during its insertion phase. An evaluation of Rayleigh wave velocities at this site yields a surface layer of 10 cm to 50 cm thickness which has a Young's modulus of at least 7 MPa, with a tentative upper limit of 980 MPa. This layer has a shear modulus in the lower third of the variability range of terrestrial snow, while its density possibly exceeds that of pore-free water ice. Both can be achieved by introducing a regolith component of loose grains. Underneath this layer, shear wave velocity is significantly reduced (32\% or more), likely due to a further reduced shear modulus \citep{Knapmeyer17}. This reduction might be the result of an incomplete sintering of the ice (and thus the absence of cementing of the non-ice components). 
Although there is no direct relation between elastic moduli and strength, some laboratory data on snow and porous ice/dust mixtures can be used to correlate Young's modulus with uniaxial compressive strength. \cite{JessbergerKotthaus89} showed that for porous ice/dust mixtures (KOSI ice) and temperatures down to 123 K the elastic modulus is about two orders of magnitude higher than the compressive strength. Assuming that this relation is also applicable  at Abydos, the compressive strength of the upper hard layer would be between 70 kPa and 9 MPa. This range is compatible with the lower limit of 2 MPa for the compressive strength deduced by \cite{Spohn15} from the fact that the MUPUS PEN was not able to penetrate into the ground at Abydos. This is also consistent with the results from experiments and modelling above (bearing in mind that compressive strength is generally greater than tensile for relevant materials).

Finally, differences between the results of the SESAME permittivity probe and the CONSERT Rosetta-to-Philae radiowave experiment can be related to an increase in the porosity (i.e. volume fraction of vacuum) from the near-surface of the small lobe of the nucleus sounded by SESAME-PP to the deep material sounded by CONSERT \citep{Kofman15, Brouet16}. These results further support the presence of a hard layer at the location of Philae.

Together, these results point to a hard sintered or recondensed layer within tens of centimetres of the cometary surface and extending down of the order of a metre, with tensile and compressive strengths around a MPa and a Young's modulus several orders of magnitude higher. Compared to this, loosely bonded granular ice should have a significantly lower tensile strength, of the  order of hundreds of kPa down to $\sim1$ kPa \citep{Kossacki15}.

\section{Modelling thermal stresses}
\label{modelling}

\subsection{Thermal model}

We computed the temperature inside the nucleus as a function of time to derive the seasonal temperature trends. We used a spherical shape model for the nucleus, orientated accordingly to its pole with RA = 69.57\,deg and DEC = 64.01\,deg and a rotational period of P = 12.40\,hr \citep{Jorda2016}. Our thermal model takes into account the solar insulation, the thermal emission, and the sublimation of water ice on the nucleus surface, and the heat conductivity inside it \citep{GroussinLamy}.  We computed the temperature on the surface and inside the nucleus over one complete revolution, taking into account the diurnal and seasonal changes in insolation with heliocentric distance. To ensure convergence, we used a time step of 12.4\,s and ran the thermal model over five complete revolutions, for a total of 8.1 millions time steps. We then computed, for each nucleus rotation (every 12.40\,hr), the average diurnal temperature at each depth interval, i.e. the seasonal trend. Our model has 2000 depth intervals of one-fifth of a diurnal skin depth, given by $\xi = I / c\rho \sqrt{P/\pi}$, where $I$ is thermal inertia in units of J\,m$^{-2}$\,K$^{-1}$\,s$^{-1/2}$, and $c=1000$\,J\,K$^{-1}$ and $\rho=532$\,kg\,m$^{-3}$ \citep{Jorda2016} are taken as the heat capacity and density of the cometary material. Overall, this allows us to examine the stress profile down to a total depth of between $\sim$4 and $\sim$20\,m, depending on the chosen value of I between 10 and \textbf{1000}\,J\,m$^{-2}$\,K$^{-1}$\,s$^{-1/2}$ (Sect.\,4.1 and Sect.\,4.2). 

With the above thermal model, we could not use a realistic shape model for 67P, since even a very  low-resolution shape model with a few thousand facets would require more than one month of computer time for each value of the thermal inertia. However, to overcome this numerical issue, we used a second thermal model, which takes into account the real shape of 67P and associated projected shadows and self-heating, but neglects thermal inertia (I = 0 J\,m$^{-2}$\,K$^{-1}$\,s$^{-1/2}$). For this purpose, we used a decimated version of shape model SHAP7 \citep{Preusker17} with 6000 facets. For each facet, we computed the temperature on the surface of the nucleus over one complete revolution, taking into account the diurnal and seasonal changes in insulation. We then computed, for each facet, the average diurnal temperature, i.e. the seasonal trend. Because of the null thermal inertia, we can only calculate stress in an infinitesimally thin surface layer and the absolute values derived from here are unrealistic. This is a conceptual limitation of the model;  nonetheless, relative stresses between facets should still reflect real differences in seasonal conditions and will be used to study variations in stress across the nucleus surface (Sect.\,4.3).

\subsection{Stress model}

We follow the method of \cite{Mellon97}, who modelled the formation of polygons in Martian permafrost. This represents an advancement over the semi-analytical approach of \cite{Kuehrt84} and \cite{TauberKuehrt87}, who used fixed rather than temperature varying parameters and only considered pure water ice and a single effective temperature profile. As described in \cite{Mellon97} ice, rocks, and frozen soils respond to applied stresses in a Maxwellian, viscoelastic way, i.e.~with elastic and viscous deformations acting in series (see their Figure 2 for a schematic representation). The total time-dependent strain can then be found by summing these two terms and, with the addition of a third term describing the thermal expansion and contraction, a thermo-viscoelastic model of the material is developed. Differentiating by time then leads to an expression for the total strain rate, $\dot{\epsilon}$, as
\begin{equation}
\dot{\epsilon} = \dot{\epsilon}_{e} + \dot{\epsilon}_{T} + \dot{\epsilon}_{v},
\label{strain}
\end{equation}
with the subscripts $e$, $T$, and $v$ referring to the elastic, thermal, and viscous terms, respectively.  Next, the assumption of a plain geometry extending in all horizontal directions is made. In this geometry the total horizontal strain rate must sum to zero as the material is restricted from expanding or contracting in this direction, and a resultant horizontal stress is induced. Using the standard elastic and thermal expressions and a creep model for the third term, as in \cite{Mellon97}, and ignoring the second-order terms, we come to the  equation
\begin{equation}
\dot{\epsilon} = 0 = \frac{1-\nu}{E}\dot{\sigma} + \alpha\dot{T} + \text{sign}({\sigma})A_{0}e^{-Q/RT}\left(\frac{\sigma}{2}\right)^n,
\label{stress}
\end{equation}
with the three terms, elastic, thermal, and viscous, as above. Here, $\sigma$ is the resultant stress in the horizontal direction; $\nu$ is the material's Poisson's ratio; $E$ and $\alpha$ are its Young's modulus and coefficient of thermal expansion; $T$ is the temperature;  $A_{0}$, $Q$, and $n$ are viscous material properties; and $R=8.31$ $\text{J  Kg}^{-1} \text{mol}^{-1}$ is the gas constant. The sign function in the viscous term is an addition from \cite{Maloof02} and ensures than the term acts in the correct direction, independent of the chosen $n$ exponent. The convention adopted here is the same of that of \cite{Mellon97}: positive values of $\sigma$ indicate tension, while negative values indicate compressive stresses.

\begin{figure}
\resizebox{\hsize}{!}{\includegraphics{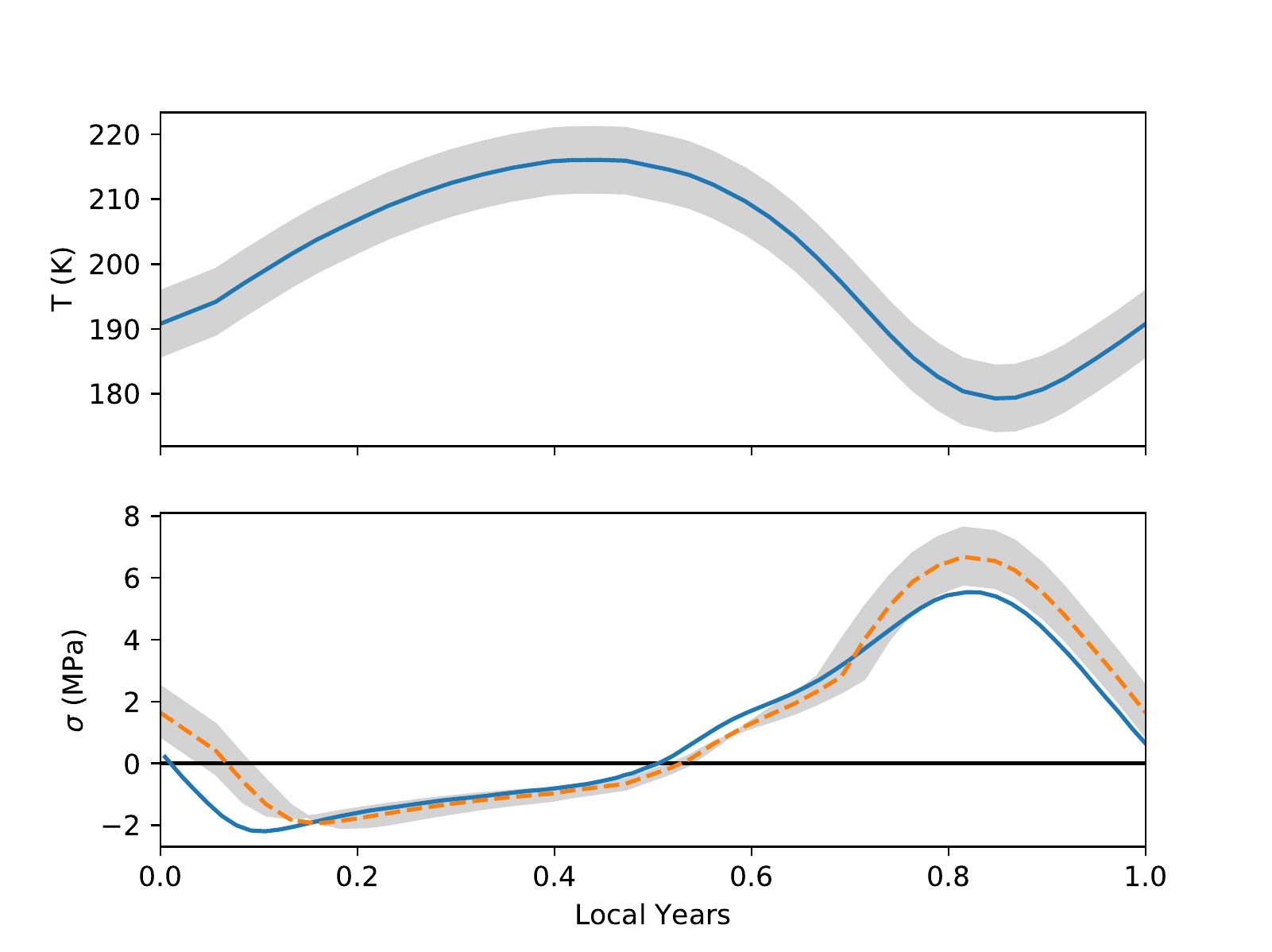}}
\caption{Reproduction of the results of Fig.~3 from \cite{Mellon97}. Top panel: Temperature over a Martian year for the Viking 2 landing site at a depth of $0.5$ m. The solid curve is a digitisation of the \cite{Mellon97} data with the grey shading representing a $\pm 5$ K uncertainty. Bottom panel: Induced thermal stresses from \cite{Mellon97} (solid) and from our code (dashed), including the uncertainty from the temperature digitalisation. RMS residuals between the \cite{Mellon97} results and ours are 746 kPa.}
\label{Fig3}
\end{figure}

Equation \ref{stress} then represents the non-linear material response to a time-varying temperature profile, and must be solved numerically for $\sigma$. When this tensile stress exceeds the material's tensile strength, fracturing will occur. We solve the equation using scientific Python's $fsolve$ routine for each time- and depth-step of a one-dimensional temperature profile. Before examining the stresses on 67P, however, we first confirm the validity of our numerical solutions by reproducing the results of \cite{Mellon97}. Figure \ref{Fig3} shows the temperature and stress at the Viking 2 landing site on Mars at a depth of $0.5$ m, from their Fig.~3. We use this temperature data as an input to our own code and, using identical material properties, produce the dashed stress curve shown in the lower panel. The overall shape and intensity of the stress is well reproduced; compressive (negative) stresses are encountered during summer as temperatures rise, while tensile (positive) stresses dominate with the falling winter temperatures. Small differences between the curves (RMS residuals of 746 kPa) may be due to a difference in time-step, which is unspecified in \cite{Mellon97}.

Having validated the code, we now run it for a seasonal cycle of the comet thermal model. This is repeated a number of times (up to several hundred) to ensure convergence. At large depths, where temperature variations are very slow, the code can still fail to fully converge. However, the trend here is for ever decreasing stress with time; furthermore,  we check each run carefully to make sure that the depths we are interested in, where the stresses change from tensile to compressive, are not affected.

For our baseline mode, we adopt the material parameters used by \cite{Mellon97} for Martian permafrost. These are for a linear mixture of ice and rock with $45\%$ ice volume, and are shown in Table \ref{tabmaterial}. Comparing the values in  Table \ref{tabmaterial} to those in the discussion in Section \ref{evidence}, it can be seen that our Young's modulus value (with range $\sim9-20$ GPa) is higher than the upper limits suggested by Philae. The measured material properties of the comet's surface are not well constrained, however, so we choose our frozen soil parameters partly for ease of comparison with previous studies. We are also justified by the fact that, as noted in \cite{Mellon97} and references therein, ice, snow (i.e.~low-density granular ice) and ice-bonded soils  have similar rheologies. Any hard layer containing significant amounts of ice should therefore exhibit the same type of viscoelastic behaviour. The stratigraphy of the comet is also poorly constrained, so we use a simple one-layer model in order to quantify the maximum induced stresses. We vary each of the main parameters in Sect.~4.4 to investigate the sensitivity of the results to these assumptions. In terms of thermal inertia, the Rosetta results imply low bulk values of  I $= 10 - 50$ J m$^{-2}$ K$^{-1}$ s$^{-1/2}$ \citep{Gulkis2015}, but a hard ice-enhanced layer may have higher thermal inertia, closer to that of pure water-ice (2000 J m$^{-2}$ K$^{-1}$ s$^{-1/2}$; \citealp{Ferrari}). We therefore run our model with values of I $= 10, 50, 200$, and 1000 J m$^{-2}$ K$^{-1}$ s$^{-1/2}$.

\begin{table*}
\begin{center}
\begin{tabular}{ccccc}
Material Property & \multicolumn{3}{c}{Pure Ice} & Mixture (soil fraction $=\phi$ ) \\
 \hline
E (Pa) & \multicolumn{3}{c}{$2.339\times10^{10} - 6.48\times10^{7} T$} & Same  \\
$\nu$ & \multicolumn{3}{c}{$0.4$} & Same  \\
$\alpha$ & \multicolumn{3}{c}{$2.47\times10^{-7}T - 1.17\times10^{-5}$}  & Linear combination with $\alpha_{soil}(273 K)=7\times10^{-6}$ \\
 \hline
& $T\ge234$ & $234>T\ge195$ & $T<195$ & \\
Q (J mol$^{-1}$) & $9.1\times10^{4}$ & $6.1\times10^{4}$ & $3.6\times10^{4}$ & Same \\
n & 4.0 & 4.0 & 4.7 & Same \\
A$_{0}$ (Pa$^{-n}$s$^{-1}$) & $10^{-12.1}$ & $10^{-18.9}$ & $10^{-31.0}$ & $A=A_{0}e^{-2n\phi}$  \\
\end{tabular}
\caption{Material properties used in our thermo-mechanical model based on those of \cite{Mellon97} and \cite{Mellon08} (see discussion and references therein for details). $E$ is the Young's modulus, $\nu$ is Poisson's ratio, $\alpha$ the coefficient of thermal expansion, $Q$ is activation energy, and $n$ and $A_{0}$ are the two viscous terms.}
\label{tabmaterial}
\end{center}
\end{table*}

\section{Modelling results}
\label{results}

\subsection{Temperature maps}

\begin{figure}
\resizebox{\hsize}{!}{\includegraphics{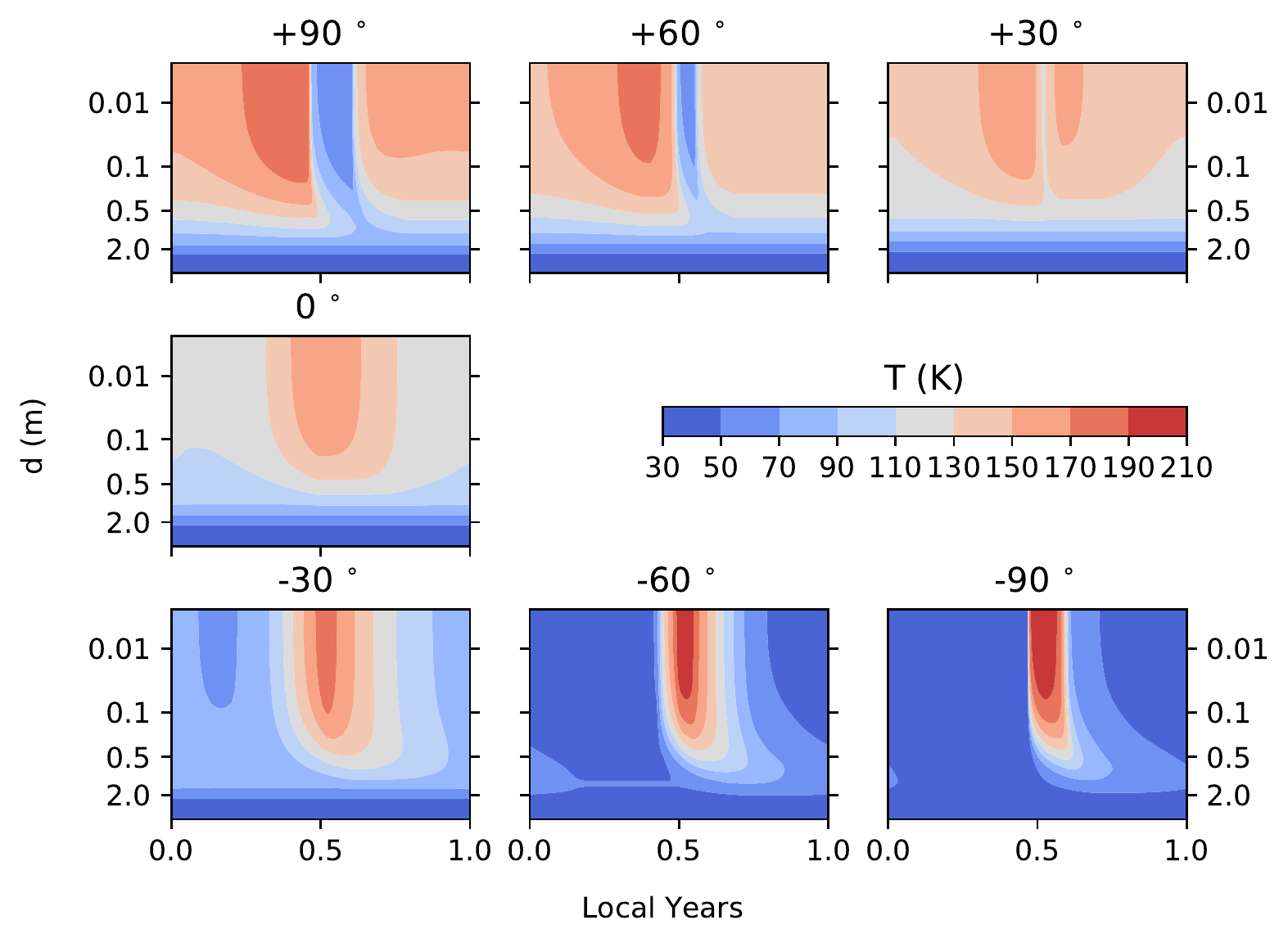}}
\caption{Temperature maps with depth and time for each of the labelled latitudes on comet 67P, produced by our 1-D thermal model with a thermal inertia of I = 50 J m$^{-2}$ K$^{-1}$ s$^{-1/2}$. Zero/one years is aphelion and 0.5 years is perihelion.}
\label{TMaps}
\end{figure}

The output of the spherical nucleus thermal model is shown for I = 50 J m$^{-2}$ K$^{-1}$ s$^{-1/2}$ in Fig.~\ref{TMaps} as maps of temperature with time and depth for each of the latitudes examined. The seasonal thermal wave can be seen propagating downwards from the surface to some depth, below which temperature variations are negligible and the material remains near the initial temperature of $30$ K.  The phasing of the seasonal change varies between the northern and southern hemispheres because of the comet's obliquity. The spin axis is aligned such that at aphelion the northern pole points roughly towards the sun and the southern pole away from it, and {vice versa} at perihelion. Therefore, as the comet approaches the Sun and insolation increases, northern latitudes become progressively warmer before experiencing a sudden temperature drop as they are orientated away from the Sun (winter), whereas southern latitudes receive very little sunshine until a sudden burst at closest approach, leading to a sharp temperature peak (summer). Because of this, surface temperatures in the southern hemisphere reach greater maximum values of just over 200 K, while peak temperatures in the north are roughly $190$ K, and less near the equator. Despite these differences, the depth to which the seasonal thermal wave penetrates is similar at all latitudes: $\sim100$ $\xi$, or $0.5-1$ m for I = 50 J m$^{-2}$ K$^{-1}$ s$^{-1/2}$. Similar results are also seen when varying I, but with correspondingly different magnitudes of depth, as discussed below.

\subsection{Stress model results}

\begin{figure}
\resizebox{\hsize}{!}{\includegraphics{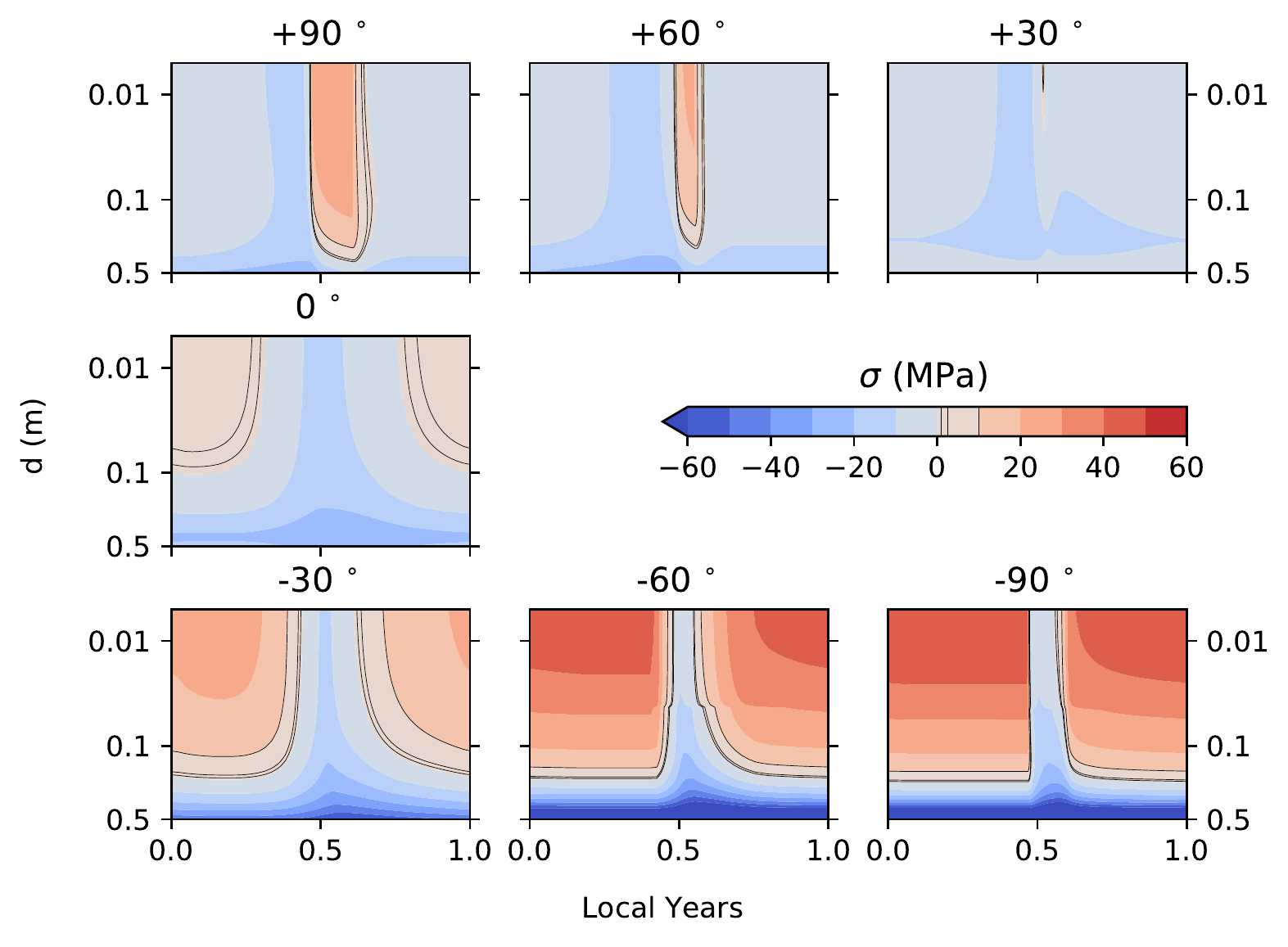}}
\caption{Thermal stresses with depth and time for each of the labelled latitudes on comet 67P, with I = 50 J m$^{-2}$ K$^{-1}$ s$^{-1/2}$. Highlighted contours are likely tensile strengths, from \cite{Kossacki15}, of $1, 2.5,$ and $10$ MPa. Zero/one years is aphelion and 0.5 years is perihelion.}
\label{Maps}
\end{figure}

Figure \ref{Maps} shows maps of the thermal stresses induced by the temperatures of Fig.~\ref{TMaps}. The general trend is for compressive stresses to follow the increasing progression of temperature in summer, followed by tensile stresses as the material cools. Variations of magnitude and timing can be seen with latitude. Tensile stresses remain high  (up to tens of MPa) for much of the cycle, particularly in the cold southern hemisphere, where the depth of the transition from tension to compression is roughly the same in all cases at $\sim0.25$ m. In the north, this depth is greater at the pole, around 0.45 m, and decreases towards the equator to only 0.1 m or less. Below around 0.5 m the model failed to converge, due to a combination of small temperature variation and high stiffness (E), and we exclude these depths from the plot.

The highlighted contours show the $1, 2.5$, and $10$ MPa values for the tensile strength of an ice-bonded hard layer from \cite{Kossacki15}. Therefore in our baseline case, thermal stresses are high enough to exceed the tensile strength of water-ice and ice-rock mixtures, and cause fracturing to depths of tens of centimetres over most of the cometary surface. In low northern latitudes the situation is less clear cut and fracturing may occur to at least between one and ten centimetres for weak materials (with tensile strengths below one MPa).

\begin{figure}
\resizebox{\hsize}{!}{\includegraphics{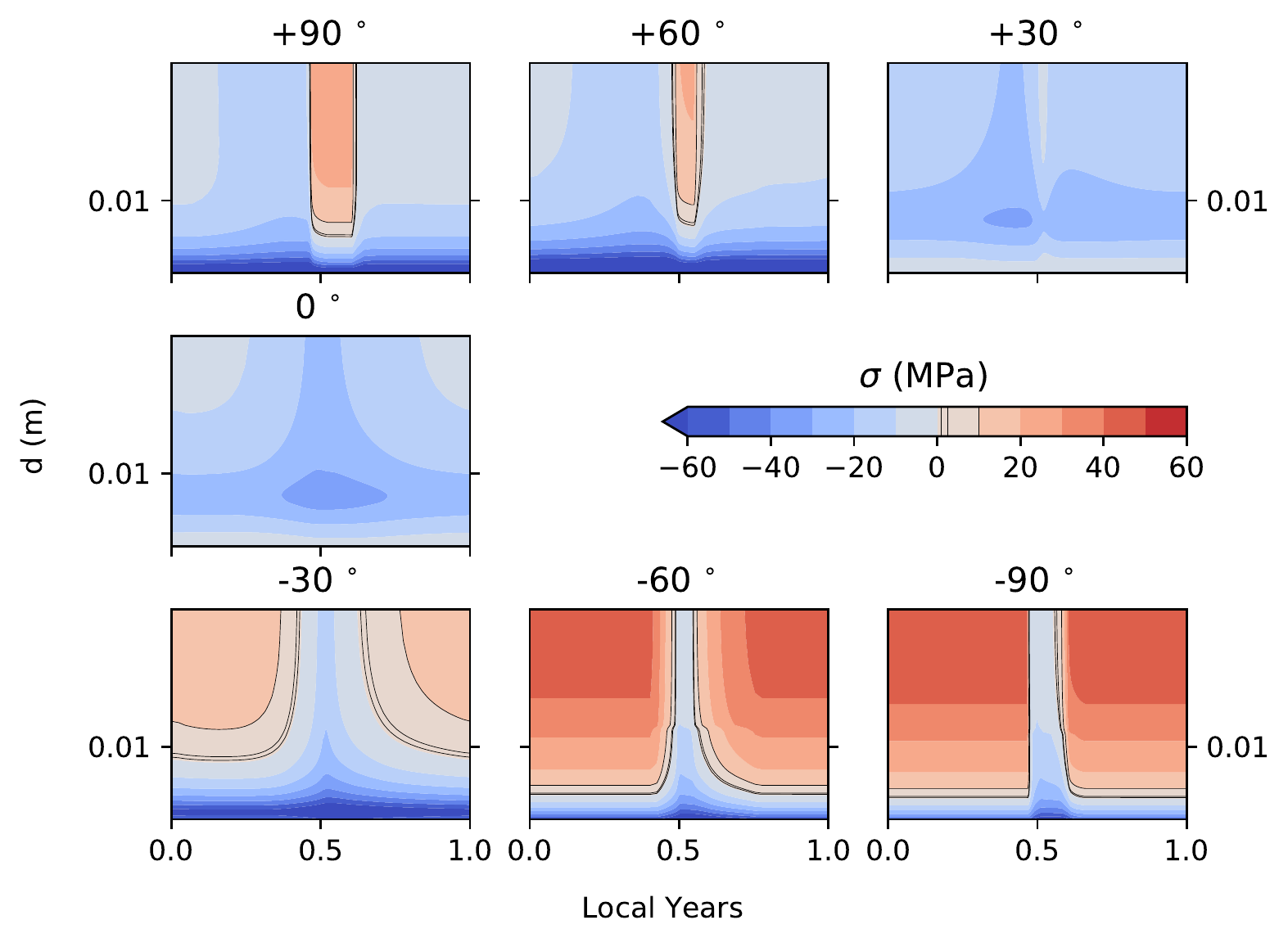}}
\caption{Thermal stresses. Identical to Fig.~\ref{Maps}, but with a thermal inertia of I = 10 J m$^{-2}$ K$^{-1}$ s$^{-1/2}$.}
\label{Maps_I10}
\end{figure}

\begin{figure}
\resizebox{\hsize}{!}{\includegraphics{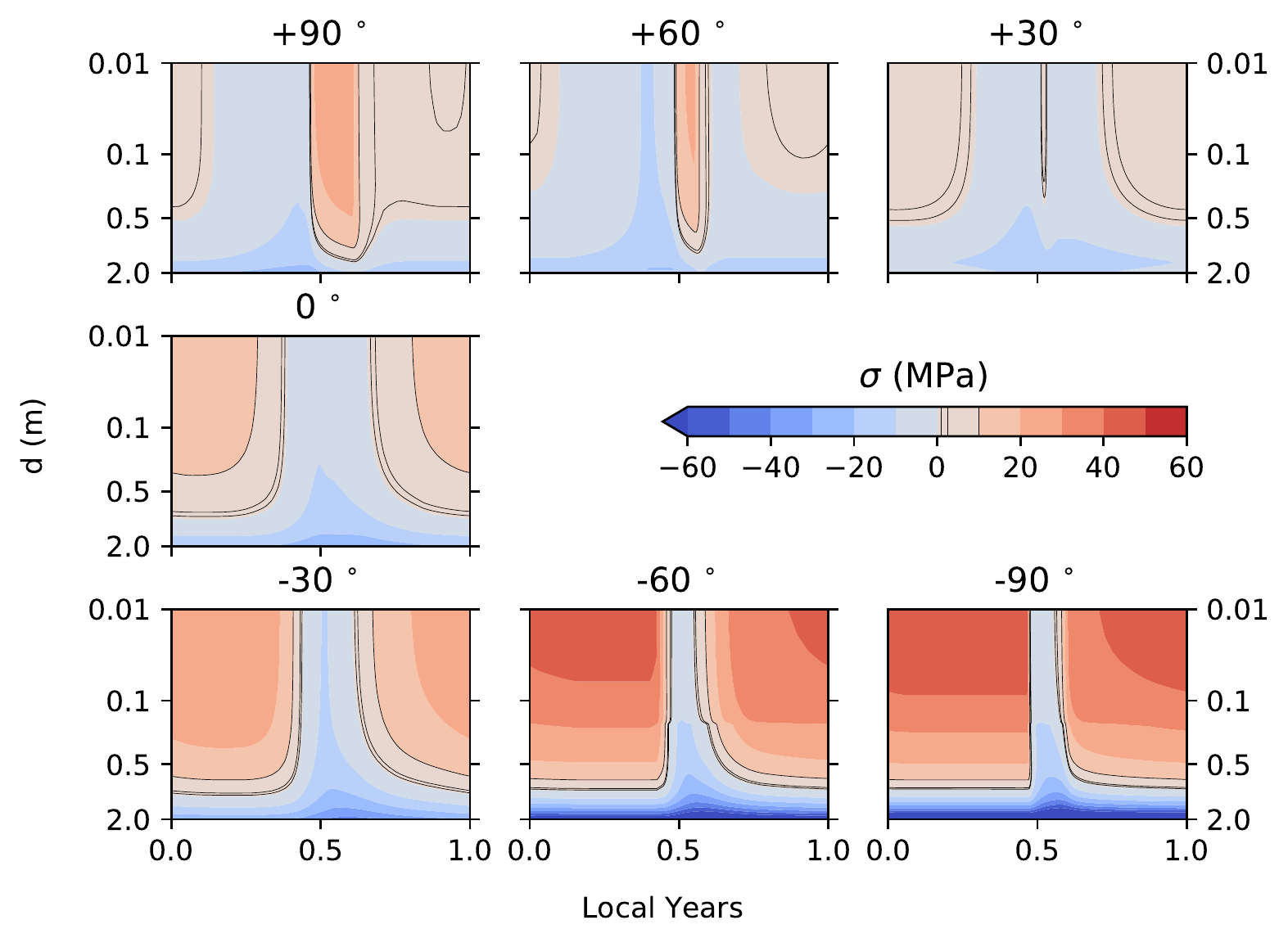}}
\caption{Thermal stresses. Identical to Fig.~\ref{Maps}, but with a thermal inertia of I = 200 J m$^{-2}$ K$^{-1}$ s$^{-1/2}$.}
\label{Maps_I200}
\end{figure}

\begin{figure}
\resizebox{\hsize}{!}{\includegraphics{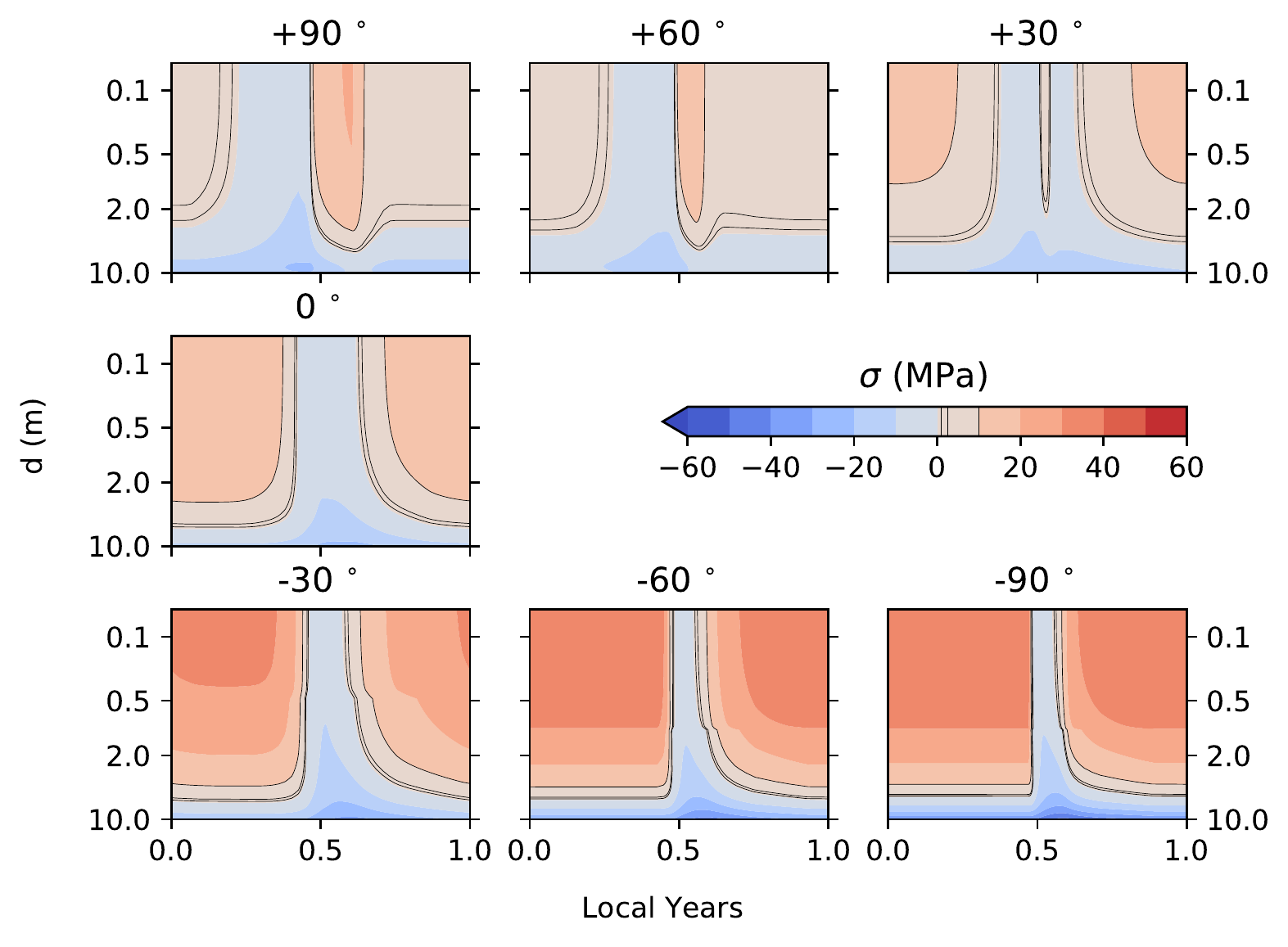}}
\caption{Thermal stresses. Identical to Fig.~\ref{Maps}, but with a thermal inertia of I = 1000 J m$^{-2}$ K$^{-1}$ s$^{-1/2}$.}
\label{Maps_I1000}
\end{figure}

Figures \ref{Maps_I10}, 5, and \ref{Maps_I1000} show the results for thermal inertia values of I = 10, 200, and 1000 J m$^{-2}$ K$^{-1}$ s$^{-1/2}$, respectively. The shapes of the stress patterns are similar in each case, but shifted up or down in depth by the differing diurnal skin depths. For I = 10 J m$^{-2}$ K$^{-1}$ s$^{-1/2}$, tensile stresses are only high enough to overcome the likely tensile strengths down to a few centimetres, and not at all near the equator, but for I = 200 J m$^{-2}$ K$^{-1}$ s$^{-1/2}$ fracturing can be expected down to at least a metre in most cases and to at least several metres for  I = 1000 J m$^{-2}$ K$^{-1}$ s$^{-1/2}$.

\begin{figure}
\resizebox{\hsize}{!}{\includegraphics{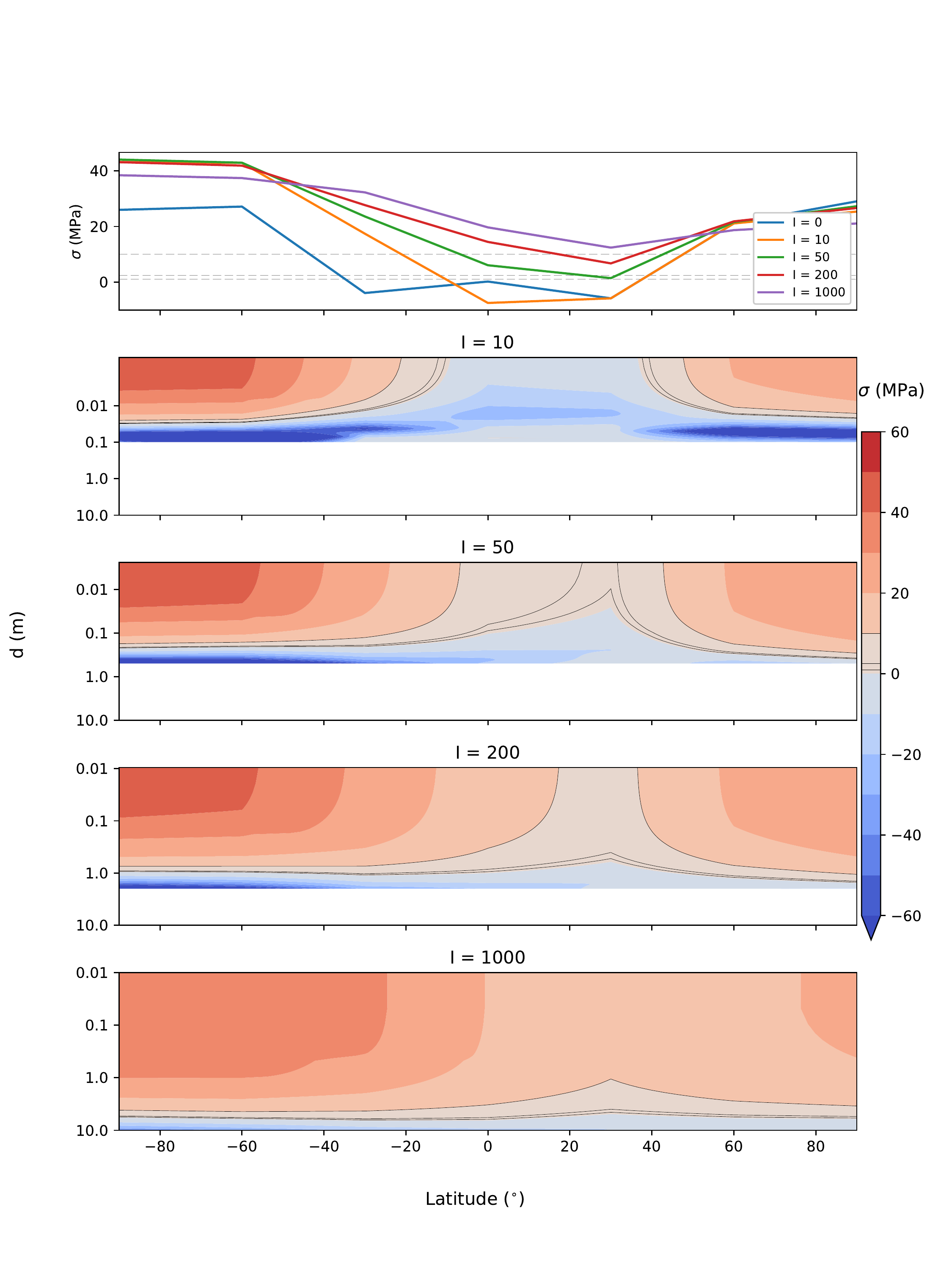}}
\caption{Top: Maximum surface thermal stress over one revolution for different thermal inertia values. Bottom: Maximum thermal stress maps with depth and latitude over one revolution for the four non-zero thermal inertia values.}
\label{Latitudes}
\end{figure}

These results are summarised in figure \ref{Latitudes}, which shows the maximum seasonal stress as a function of latitude for the three thermal inertia values. The highlighted contours again show values of tensile strength from \cite{Kossacki15}. The top panel shows the maximum stress encountered at the surface for the three values of thermal inertia, plus an additional curve computed using no thermal inertia, I = 0 J m$^{-2}$ K$^{-1}$ s$^{-1/2}$. The band of low stress centred just north of the equator is clearly seen in all cases, with a width of between around 80 and 20 degrees depending on I. This represents the region with the narrowest seasonal temperature range (as shown in Fig.~\ref{TMaps}) due to the particular orientation of the comet's obliquity. A trend for increasing stress with thermal inertia can also be seen near the equator; this results from the higher average diurnal temperatures produced by higher thermal inertia during the summer. The results of the zero thermal inertia model show stresses that are everywhere equal to or lower than the I = $50-1000$ J m$^{-2}$ K$^{-1}$ s$^{-1/2}$ curves, and generally lower than the I = 10 J m$^{-2}$ K$^{-1}$ s$^{-1/2}$ curve, except at the equator itself. A model without thermal inertia can therefore be considered as a lower estimate of thermal stress.

\subsection{Stress results on the shape model}

\begin{figure}
\resizebox{\hsize}{!}{\includegraphics{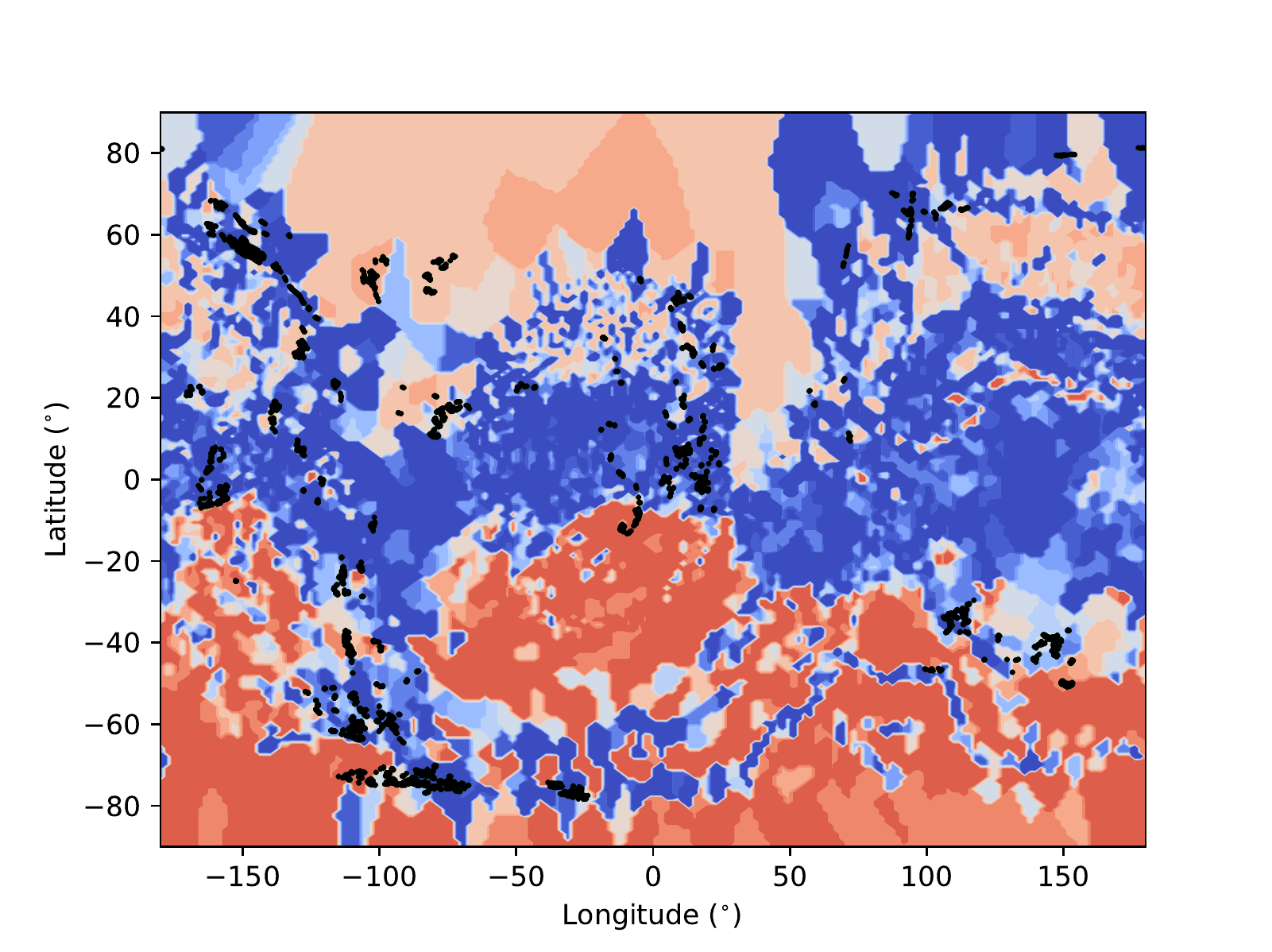}}
\caption{Maximum thermal stresses on the surface of the 67P shape model for the I = 0 J m$^{-2}$ K$^{-1}$ s$^{-1/2}$ model. Values are for qualitative comparison only. Black points are the locations of the thermal contraction crack polygons measured in \cite{Auger17}. No significant differences are noted between the distribution of stresses on all facets and those containing polygons.}
\label{stress_map}
\end{figure}

We compare these results to the simplified thermal stress model using the comet shape, and to the observed locations of polygons. Figure \ref{stress_map} shows the maximum seasonal stress found at the surface using this approach. Thermal stresses broadly follow the pattern of Fig.~\ref{Latitudes}, with the greatest stresses near the south pole, intermediate stress near the north pole, and the lowest values near the equator and in the low-latitude northern regions. Significant `patchiness' is seen because of the effects of local topography and the low resolution of the shape model used. Absolute stress values are not shown here due to the lack of thermal inertia making them unrealistic; the figure is mainly presented to compare stresses in different regions in a qualitative manner. As described in the previous section, however, this zero thermal inertia model can be considered as a minimum stress estimate compared to models with realistic thermal inertia values.

Also shown in Fig.~\ref{stress_map} are the locations of the polygons measured by \cite{Auger17}. They are found all over the comet, at varying latitudes and on facets with low and high stress, with no apparent correlation to peak thermal stress. No significant differences are noted between the distribution of stress on all facets and those containing polygons. As noted in \cite{Auger17} there is a bias against the detection of polygons in the southern hemisphere, and therefore the highest stress regions, due to incomplete image coverage here at the time of the survey. The presence of polygons everywhere suggests stresses exceeding the material strength at all latitudes and, since this only occurs for the higher thermal inertia models described above, we infer I $\gtrsim 50$ J m$^{-2}$ K$^{-1}$ s$^{-1/2}$ for all the regions where polygons are observed.

\subsection{Sensitivity to material parameters}

We now examine the sensitivity of these results to changes in the material parameters, many of which are poorly constrained.

\begin{figure}
\resizebox{\hsize}{!}{\includegraphics{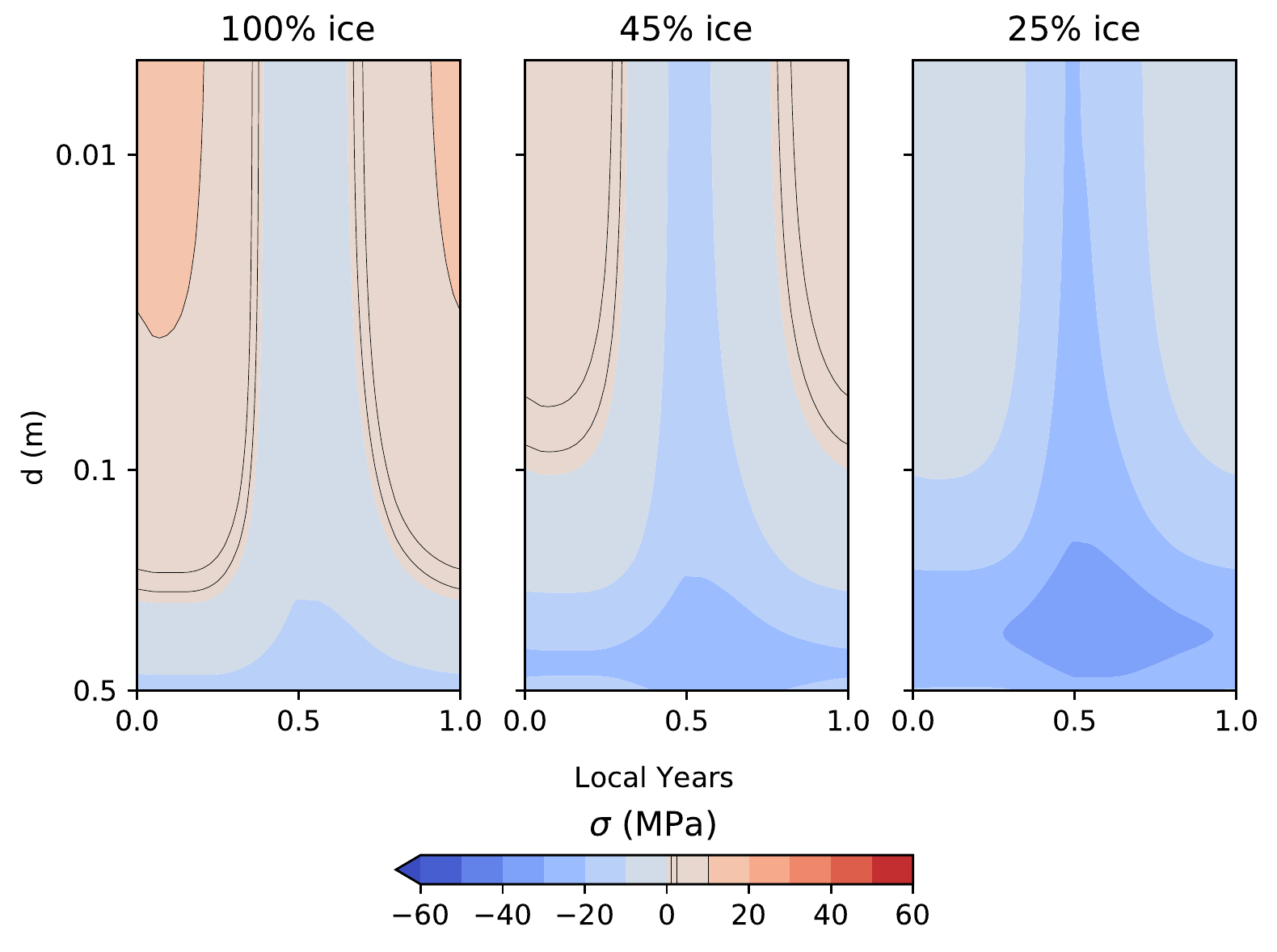}}
\caption{Thermal stress for the equator of 67P with I = 50 J m$^{-2}$ K$^{-1}$ s$^{-1/2}$ and three different values of ice volume fraction.}
\label{Icecontent}
\end{figure}

\begin{figure}
\resizebox{\hsize}{!}{\includegraphics{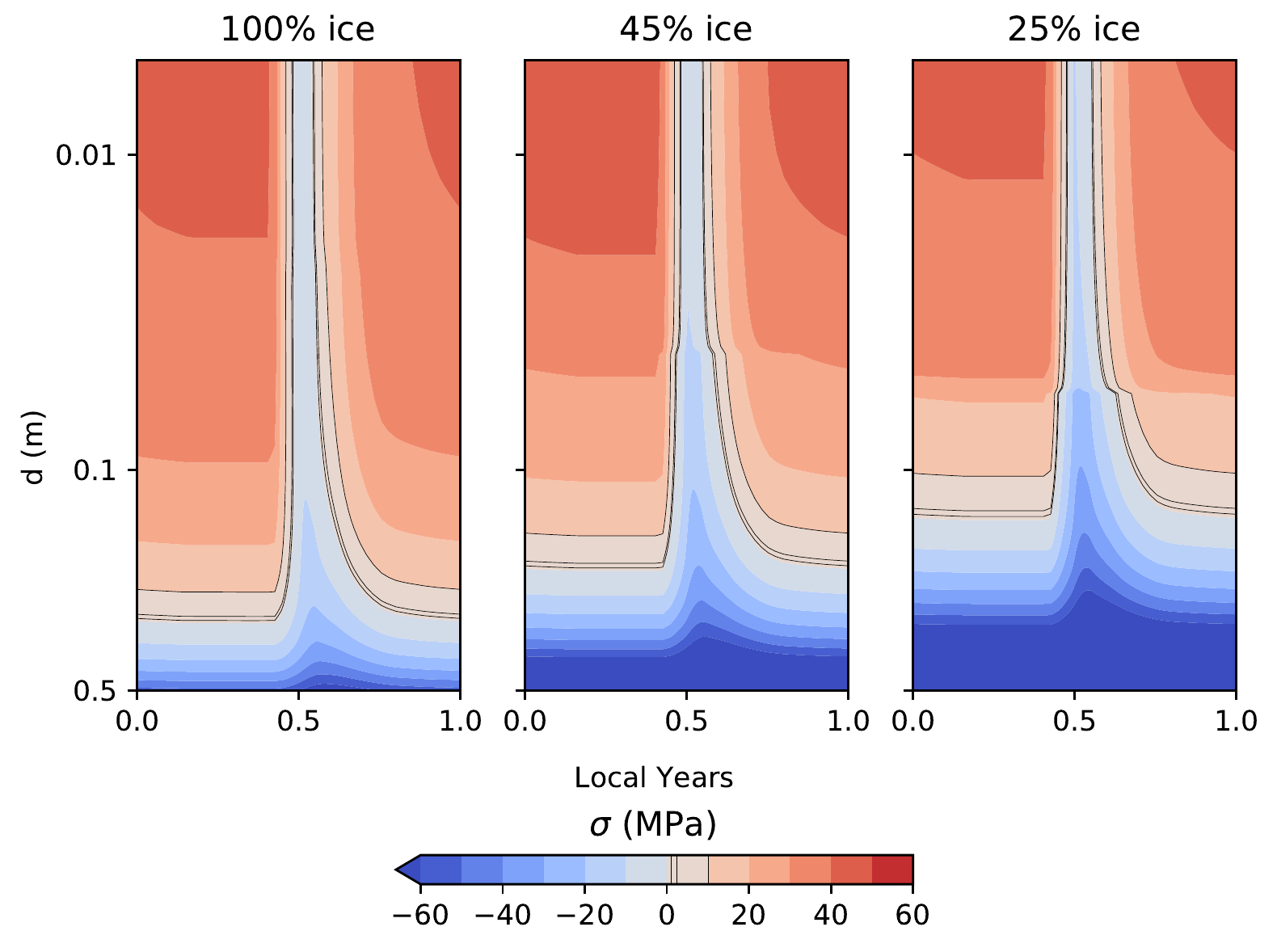}}
\caption{Thermal stress for a latitude of $-60^{\circ}$ with I = 50 J m$^{-2}$ K$^{-1}$ s$^{-1/2}$ and three different values of ice volume fraction.}
\label{icecontent_lat-60}
\end{figure}

In Fig.~\ref{Icecontent} we examine the effects of changing the ice content of the modelled ice-rock mixture from our baseline $45\%$ ice volume, using the equatorial I = 50 J m$^{-2}$ K$^{-1}$ s$^{-1/2}$ case as an example. This is achieved by modifying the thermal model as well as the thermal expansion parameter, $\alpha$, and the viscous term, as shown in Table \ref{tabmaterial} and described in \cite{Mellon97} and \cite{Mellon08}. A trend for decreasing stress with ice content is seen, with no fracturing predicted to occur for the $25\%$ ice volume case. For even lower ice content, thermal expansion is small and the model begins to break down and experience convergence problems at smaller and smaller depths. Figure \ref{icecontent_lat-60} shows the same test for a latitude of $-60^{\circ}$. The same trend is observed, but here the stresses are high enough that fracturing should still occur even with small amounts of ice.

The trends in these plots are driven by the large thermal expansion coefficient of ice. Holding the other parameters constant and varying $\alpha$ on its own produces a similar effect, a roughly linear relation between peak stress and thermal expansion, while varying the viscous constant, $A_{0}$, on its own has little effect. Similarly, the viscous exponent, $n$, appears to have little effect when varied between the different values of \cite{Mellon97} and \cite{Mellon08}. This is consistent with the results of \cite{TauberKuehrt87}, who suggested viscous effects to be of lesser importance than elastic ones. Varying ice content will, additionally, change the Young's modulus and tensile strength. As shown in the contours in Figs.~\ref{Maps} to \ref{Latitudes} above, tensile strength is two to three times higher for ice-rock mixtures than for pure water ice \citep{Kossacki15}. \cite{Mellon97} describe the baseline Young's modulus used here as conservative, and note that coarse-grained mixtures are found to have values up to five times higher than pure ice, whereas Philae's measurements of the hard layer suggest a value an order of magnitude lower. Holding the other parameters constant and varying $E$ by an order of magnitude in either direction produced roughly corresponding changes in peak stress, suggesting that the two effects are approximately balanced (i.e.~greater rock content will decrease the Young's modulus, and therefore the induced stresses, but also reduce tensile strength by the same order of magnitude), leaving the thermal expansion coefficient to once again dominate.

From this we can see that the primary influences on thermal stress are thermal inertia, ice content, and the material strength and stiffness (tensile strength and Young's modulus) of the hard layer. If thermal inertia varies across the comet then the depths to which high stresses penetrate will be different in different regions. This would affect the depths of crack propagation and distance between adjacent fractures. The other mechanical parameters are interconnected, so that a strong, stiff material may be more resistant to thermal expansion, while a weaker material (lower $E$) will experience lower stress, but will also require lower stresses to fracture (lower tensile strength). Unfortunately, the properties of partially bonded granular materials and the relations between these parameters are poorly understood. However, we can say that a hard layer with thermal inertia $\gtrsim 50$ J m$^{-2}$ K$^{-1}$ s$^{-1/2}$, and with significant quantities of water ice ($\gtrsim 45\%$ at the equator, less elsewhere) should fracture, whatever its locations on the comet. If no such layer exists, and the material is purely unconsolidated and granular, then it is difficult to see how fracturing can be supported with our model.

\section{Discussion and conclusions}
\label{conclusion}
\subsection{Summary}
The results of our thermo-viscoelastic model show stresses of up to several tens of MPa induced by seasonal temperature changes on a spherical model of comet 67P down to depths of between a centimetre and a metre. These are seen at most locations on the comet, with the exception of a band of low stress in the middle and low latitudes of the northern hemisphere (Fig.~\ref{Latitudes}), which is particularly pronounced for low values of thermal inertia (I $\lesssim10$ J m$^{-2}$ K$^{-1}$ s$^{-1/2}$). Introducing local topography, in the form of 67P's complicated shape (but neglecting thermal inertia), limits these regions of low surface stress to patchy areas, mostly still in the north and equatorial region (Fig.~\ref{stress_map}).
On the spherical model, stress increases with increasing thermal inertia and, for values of I = $50 - 1000$ J m$^{-2}$ K$^{-1}$ s$^{-1/2}$ (and ice volume fractions $\gtrsim 45\%$), exceeds reasonable material tensile strengths of $\sim1$ MPa at all latitudes, leading to fracturing.

Additionally, stresses vary with ice content, Young's modulus, and thermal expansion coefficient. If these material parameters, as well as thermal inertia, vary across the surface then fracturing will not be uniform. Stress decreases with decreasing ice content and at zero ice-fraction the unconsolidated material is probably not competent enough to support fractures. Between this and the lower end of our simulations ($25\%$ ice) the minimum ice content for fracturing is hard to quantify, and varies with thermal inertia and latitude (see Figs.~\ref{Icecontent} and \ref{icecontent_lat-60}). The strongest constraints are found in the equatorial region, where $45\%$ ice is required. For comparison, \cite{Mellon97} note that soils with low ice content behave similarly to dry soil (and ice with small impurities retains ice-like rheology), while thermal contraction polygons have been observed in terrestrial surfaces ranging from nearly pure ice \citep{Marchant02} to permafrost with $10\%$ or less ice content \citep{Bockheim95}. Various Rosetta measurements \citep{Fulle2016} point to a global dust-to-ice ratio for 67P's nucleus of around 6, or an average of $15\%$ water-ice, insufficient to cause fracturing at all latitudes in our model (though probably sufficient at the poles). A higher than average ice content may well be expected, however, in the hardened layer of interest. \cite{Oklay}, for example, saw ice-enhanced regions on the surface and boulders with $\sim48\%$ and $\sim25\%$ ice content, respectively, suggesting that enhancements in an ice-rich subsurface layer to the volume fractions examined here are not unreasonable.

To summarise, a hard layer, at a depth of a few centimetres to metres and with sufficient thermal inertia (I $\gtrsim 50$ J m$^{-2}$ K$^{-1}$ s$^{-1/2}$) and ice content, will undergo thermal fracturing globally. Near the poles this layer may have approximately the global average ice content, but near the equator it must be significantly more icy ($\gtrsim 45\%$) and/or have a higher thermal inertia for fracturing to occur.

\subsection{Comparison with observed thermal contraction crack polygons}
Thermal contraction polygons are found all over 67P's surface (Fig.~\ref{stress_map}). There appears to be little correlation between peak stress, as predicted by our model, and polygon location.

Wherever polygons are observed, we predict thermal inertia and ice content values of I $\gtrsim 50$ J m$^{-2}$ K$^{-1}$ s$^{-1/2}$ and $\gtrsim 45\%$ (for the equator), respectively, at the time of formation. For the polar regions, lower thermal inertia and ice content constraints are placed because of the higher temperature ranges and stresses. Higher inertias are expected on the consolidated terrains than on the dust covered surfaces, and this is indeed where polygons are found. The presence of polygons on consolidated, high thermal inertia terrains is consistent with current seasonal thermal stresses, but we  cannot also rule out ancient polygons resulting from differing thermal environments due to past orbital or obliquity changes. 

In terms of polygon size, examples on Earth and Mars typically have a distance between adjacent cracks of three \citep{Lachenbruch62} to ten \citep{ElMaarry10} times their fracture depth. Fracture propagation is complex \citep{Lachenbruch62} and not well studied in a vacuum \citep{Molaro15}, but is enhanced in cold, brittle materials under low hydrostatic pressure, as occurs on the comet. Under such conditions, fractures can extend to depths $3-15$ times that of the high-stressed region \citep{Maloof02}. From Fig.~\ref{Latitudes}, therefore, thermal fractures of 67P penetrate to between three centimetres and fifteen metres deep, with expected polygons of between 0.09 and 45 m across, but likely $\sim1-20$ m for reasonable (I = 50 J m$^{-2}$ K$^{-1}$ s$^{-1/2}$) thermal inertia values. This is consistent with the measured polygons of mean size 3 m and $90\%$ between one and five metres \citep{Auger17}. Observed polygon size is relatively uniform across the comet, suggesting little variation in the depth or mechanical properties of the hard layer. This is, however, slightly puzzling when compared to the variation in the depth of the stressed layer shown in Fig.~\ref{Latitudes}. It is possible that variations in topography and thermal environment change the stress locally (as shown in Fig.~\ref{stress_map}) confusing any smooth variation of polygon size with latitude.

\subsection{Implications for erosion}

Thermal fracturing on the metre-scale should be an important erosion mechanism. Gradual weakening of material by thermal fatigue will break down boulders into smaller and smaller pieces \citep{Pajola15}, whilst the presence of fractures in cliff walls will weaken the material, making collapse under its own weight more likely. Debris fields at the bottom of many cliffs suggest this is a common process \citep{ElMaarry15, Groussin15}, while individual collapses have been linked with outburst activity \citep{Pajola2017}. \cite{Vincent16} suggest that intact but fractured cliffs allow heat to penetrate further into the underlying volatile materials, increasing sublimation and driving activity. Our model supports and validates the idea of weakening material with thermal stresses.

\begin{figure*}
\centering
\includegraphics[width=17cm]{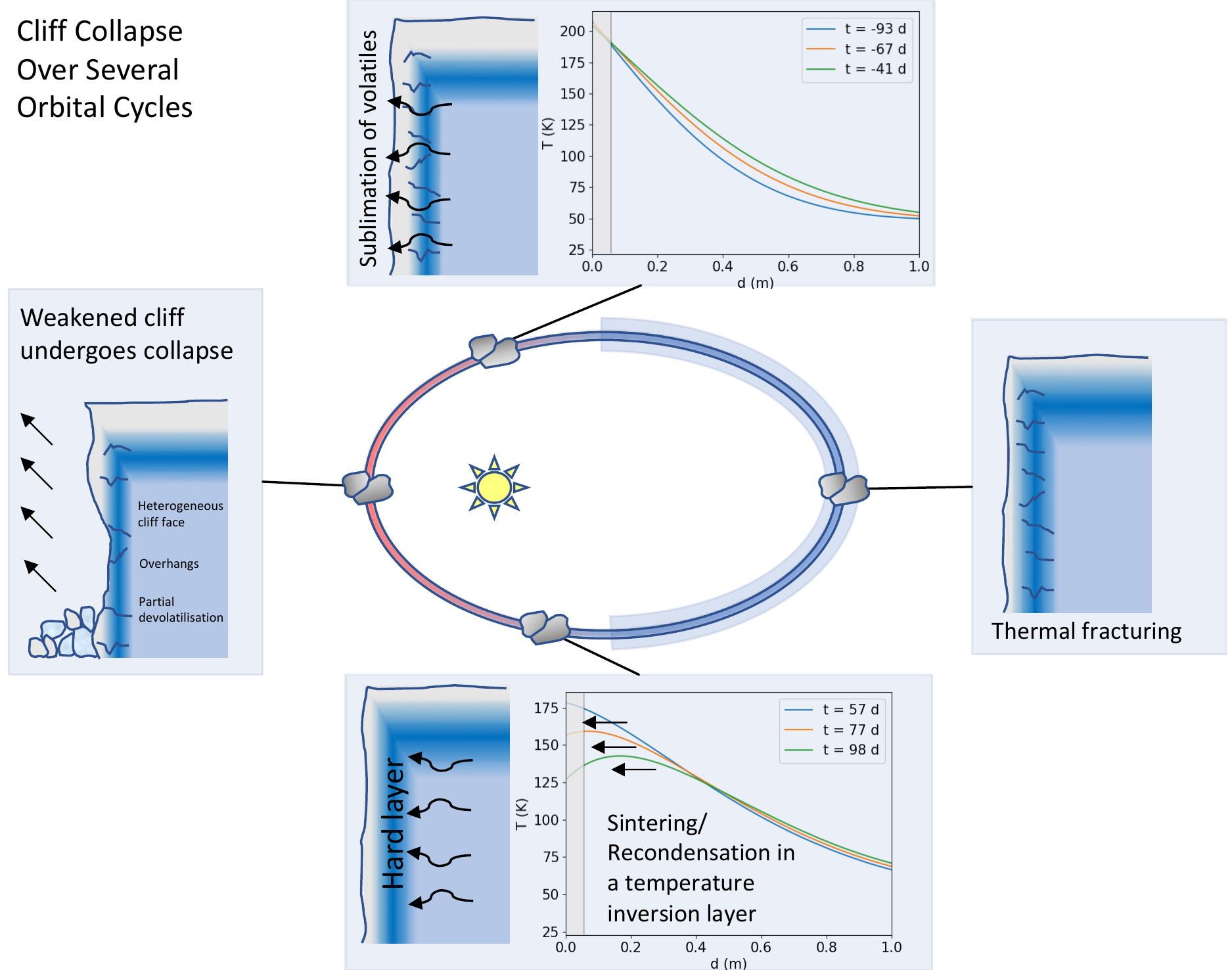}
\caption{Model for the collapse of cliffs driven by thermal stresses. Solar insolation heats exposed interior material on the face of cliffs, driving the sublimation of volatiles, whilst flat, dust covered surfaces are insulated from large temperature changes. Temperature inversions can occur as the seasonal thermal wave penetrates to depth, even as the surface begins to cool again, allowing recondensation of the volatiles in the top layers. Recondensed water-ice forms a hard layer which is vulnerable to thermal fracturing, leading eventually to cliff collapse. Temperature profiles are taken from the I $=50$ J m$^{-2}$ K$^{-1}$ s$^{-1/2}$, latitude $= -90^{\circ}$ case, for times in days since perihelion. Grey boxes show the regions, within five skin depths of the surface, where diurnal temperature changes dominate.}
\label{Cliffcartoon}
\end{figure*}

We extend these ideas of eroding cliff activity by first considering the formation and fracturing of a hard layer. Figure \ref{Cliffcartoon}, shows a cartoon of a cliff topped by a dust mantle. If the dust-layer is thick enough, then it will insulate the material immediately below it from large temperature changes, but the exposed consolidated material of the cliff wall will suffer the full seasonal temperature profile,  as shown in Fig.~\ref{TMaps}. Volatiles in the outer layers will sublimate and begin to flow outwards through the cliff face as the comet heats up. During the comet's retreat from the Sun, the cliff surface will begin to cool again which, in some cases, can lead to a temperature inversion where the surface layers are cooler than the interior. In Fig.~\ref{Cliffcartoon} we show temperature profiles extracted from our thermal model for the south polar case, where clear temperature inversions can be seen shortly after perihelion. Sublimating water ice will recondense in these regions (the upper $\sim10$ cm in this case) and may form a hard layer which will experience high stresses and thermal fracturing on the next, and subsequent, orbital cycles. Once fractured, the cliff is more vulnerable to partial collapse, leaving an overhang, full collapse into a debris field, and the sort of enhanced gas flow and activity proposed by \cite{Vincent16}. This   process could be occurring all over the comet, leading to the gradual retreat of cliffs and removal of material across the surface.

It should also be noted that the same mechanisms will apply to other objects, similar to 67P. Other Jupiter-family comets undergo comparable seasonal temperature changes and, therefore, the presence of hard, ice layers and thermal fracturing is expected on them as well. Thermal contraction polygons may well be found on other comets, if observed with sufficient resolution \citep{Auger17}, and the linked processes of fracturing, cliff erosion and activity may also be a common feature \citep{Vincent16}.
\subsection{Perspectives}
In this work we have considered two models: the simplified cases of a uniform material with depth on a spherical nucleus, and a material without thermal inertia on a shape model with complex topography. A more complete model could combine both of these to simulate temperature with depth, thermal inertia, and topography, but as noted this would be very computationally expensive.

We also did not investigate diurnal temperature changes, which are similar in magnitude to the seasonal changes, but on shorter timescales, leading to high rates of change. The limited penetration of the diurnal temperature wave (skin depth of $1-2$ cm; \citealp{Gulkis2015}), however, should limit the induced stresses to only the very top layers. Thus, diurnal stresses may well be important in the weathering of the outer centimetres of boulders and consolidated material, but are unlikely to play a role in the formation of polygons and other metre-scale fractures. Fracturing has indeed been observed at these small scales \citep{Poulet16} and may be evidence for such a scenario.
The fracture networks observed by CIVA do not resemble the regular polygons found at larger scales. However, the stress pattern may be modified by pre-existing weaknesses in the micro-structure (which can affect polygon shape; \citealp{Lachenbruch62, Auger17}) or by other stresses, such as from desiccation. Desiccation-formed fractures on Mars are similar in morphology to thermal contraction polygons \citep{ElMaarry10}, but the interactions of different stresses from seasonal and diurnal temperature changes, thermal fatigue, and thermal shock, and of desiccation, have not been well studied.

Finally, the work might be extended by following the finite element modelling of \cite{Mellon08} to further investigate the relationship between stress, ice content, and polygon size. Nevertheless, uncertainties in material properties would involve a large parameter space.

\begin{acknowledgements}
This project has received funding from the European Union's Horizon 2020 research and innovation programme under grant agreement no. 686709. This work was supported by the Swiss State Secretariat for Education, Research and Innovation (SERI) under contract number 16.0008-2. The opinions expressed and arguments employed herein do not necessarily reflect the official view of the Swiss Government.
\end{acknowledgements}

\bibliographystyle{aa}
\bibliography{Bibliography}

\end{document}